\documentclass[12pt,preprint]{aastex}

\slugcomment{Not to appear in Nonlearned J., 45.}

\shorttitle{3 poststarburst galaxies} \shortauthors{Yufeng Mao et
al.}

\usepackage{amssymb}
\usepackage{pifont}
\usepackage{amsmath}

\begin{document}


\title{Three Spectacular HII-buried-AGN Galaxies from SDSS}


\author{Yufeng Mao\altaffilmark{1,2}, Jing Wang\altaffilmark{1}, \& Jianyan Wei\altaffilmark{1}}
\altaffiltext{1}{National Astronomical Observatories, Chinese
Academy of Sciences, Beijing 100012, China; myf@bao.ac.cn}
\altaffiltext{2}{Graduate University of Chinese Academy of Sciences,
Beijing, P.R.China}


\begin{abstract}

We present our analysis of the three HII-buried-AGN: SDSS
J091053+333008, SDSS J121837+091324, and SDSS J153002-020415, by
studying their optical spectra extracted from SDSS. The location in
the BPT diagnostic diagrams of the three galaxies indicates that the
narrow emission lines are mainly exited from HII regions. However,
after the removal of the host galaxy's stellar emission, the
emission lines display the typical feature of Narrow-line Seyfert
1-like. All of the three objects have large Eddington ratio, small
black hole mass, and low star formation rate. We propose that the
three galaxies are at the transit stage from the starburst-dominated
phase to AGN-dominated phase.

\end{abstract}


\keywords{galaxies: active --- quasars: emission lines
--- quasars}



\section{Introduction}

In recent years, there have been remarkable evidences that the
evolution of supermassive black hole (SMBH) and its host galaxy are
tightly linked. The significant correlation between the mass of the
SMBH and the velocity dispersion of the bulge where the SMBH resides
in implies that there exists a tight relation between AGN activity
and star formation in the bulge (e.g. Magorrian et al. 1998; Shields
et al. 2003; Greene \& Ho 2006; Tremaine et al. 2002; Ferrarese et
al. 2006). It is now clear that AGN activity and star formation
frequently happen together (e.g. Gonzalez Delgado 2002). A specific
example is the ``Q+A'' galaxies, characterized by composite spectra
displaying broad emission lines as well as Balmer absorption lines
of A-type stars (e.g., Brotherton et al. 1999; Zhou et al. 2005;
Wang \& Wei, 2006; Goto 2006; Wild et al. 2007).

Accumulating studies show a likely evolution sequence of
co-evolution of AGN and star formation. Using the large SDSS spectra
database, Heckman et al. (2004) found that most accretion occurs
onto the black holes with high stellar surface mass densities and
young stellar populations. Wang et al. (2006) found that the
well-documented Eigenvector 1 (E1) space (Boroson \& Green, 1992;
hereafter BG92) could be extended to the infrared color
$\alpha$(60,25). They argued that AGNs might evolve along with the
E1 space. Wang \& Wei (2008) recently studied a sample of partially
obscured AGNs from SDSS. The broad H$\alpha$ emission inferred
Eddington ratio ($L/L_{\rm{Edd}}$) is found to decrease with age of
stellar population. This result is consistent with the studies of
Kewley et al. (2006) and Wild et al. (2007), who estimated
$L/L_{\mathrm{Edd}}$ indirectly through
$L(\mathrm{[OIII]})/\sigma_*$ for obscured AGNs.

However, we still know very little about the details of the
co-evolution of the starbursts and SMBH. For instance, do AGN
activity trigger starbursts (e.g. Goncalves et al. 1999), or
starbursts trigger AGN activity (e.g. Weedman 1983), or they happen
at the same time and co-evolve together? There are some suggestions
from the works on the highest redshift quasar that the black holes
form prior to the assembly of the stellar bulges (Walter et al.
2004; Riechers et al. 2008). Numerical simulations of galaxies
merger including SMBHs developed a theoretical light curve of
central AGN activity (Di Matteo et al. 2005; Springel et al. 2005).
In this model, for most of the duration of the starburst, the black
hole is ``buried'', being heavily obscured by surrounding gas and
dust, especially in UV/optical bands (Hopkins et al. 2005a,b). If
so, we would expect to identify some AGNs with their narrow emission
lines dominantly exited from \ion{H}{2} region, indicating that the
central AGNs are surrounded by the gas or dust of the \ion{H}{2}
region.

In this paper, we report the identification of three
``HII-buried-AGN'' from SDSS Data Release 4 (DR4, Adelman-McCarthy
et al. 2006) MPA/JHU catalog \footnote{The catalogs could be
downloaded from http://www/mpa-garching.mpg.de/SDSS/.}. They are
SDSS\,J091053 +333008, SDSS\,J121837+091324, and
SDSS\,J153002-020415. Their redshifts are 0.116$\pm$0.001,
0.078$\pm$0.001, and 0.051$\pm$0.001, respectively. All of the three
objects show typical AGN features, such as power-law continuum,
broad emission line of H$\alpha$ and H$\beta$, and even strong
optical \ion{Fe}{2} emissions. However, by fitting the narrow
emission lines, we can determine that all of the three objects are
located in the \ion{H}{2} regions in the BPT diagnostic diagrams
(Baldwin et al. 1981). The paper is structured as follows. We
present the objects selection in \S 2, and introduce the spectral
analysis in \S 3. The results and discussions are presented in
Section 4 and 5, respectively. The $\Lambda$ cold dark matter
($\Lambda$CDM) cosmology with $\Omega_{m}=0.3$,
$\Omega_{\Lambda}=0.7$, and ${\rm H_{0}=70~km~s^{-1}~Mpc^{-1}}$
(Spergel et al. 2003) is assumed throughout this paper.

\section{Objects selection}

Because the blueshifted [OIII] emission is thought to be an
indicator of outflows in AGN (e.g. Boroson 2005; Komossa et al.
2008), our primary motivation is to compare the different properties
of blueshifted [OIII] profile between star-forming galaxies and
Seyferts + LINERs, and results will be presented in subsequent
papers. We start our objects selection from SDSS DR4 MPA/JHU
catalogues (Kauffmann et al. 2003; Heckman et al. 2004). The
catalogs contain a set of physical properties of 567,486
emission-line galaxies, including Seyferts, LINERS, and star-forming
galaxies. To ensure our sample have a high quality in spectra, we
need our sample have a median S/N larger than 20 across the whole
spectrum. Especially, we need the [OIII] and H$\beta$ emission lines
have S/N larger than 40. About 3200 star-forming galaxies have been
selected according to the criteria. Then we focus on the 200
star-forming galaxies which show the largest [OIII] blueshift
($\sim$100 km\ $\rm{s^{-1}}$), and check their spectra by eyes
carefully. Only three objects among them, SDSS J091053+333008, SDSS
J121837+091324, and SDSS J153002-020415, display typical AGN
features, such as broad Balmer emission lines and FeII complex,
superposed on spectra of young stellar population stars. The evident
AGN features motivate us to carefully re-examine the spectra
properties for these three objects.

\section{Spectral analysis}

The spectral reduction is performed by standard IRAF procedures. By
assuming the Galactic extinction curve with $R_V=3.1$, each spectrum
is corrected for Galactic extinction correction following the
reddening maps of Schlegel et al. (1998). Each of the
extinction-corrected spectrum is transformed to the rest frame,
along with $k$-correction, according to the corresponding redshift
provided by the SDSS pipelines. The spectra at the rest frame are
displayed in Figure 1.

For SDSS\,J091053+333008, because its star light is so weak that the
spectrum is dominated by AGN's featureless continuum and broad
emission lines, we fit the continuum by the combination of a free
powerlaw and the \ion{Fe}{2} template provided by BG92. The FWHM of
the FeII template is fixed to that of its H$\beta$ broad component
(e.g., BG92; Hu et al. 2008). For the other two objects, the star
light components are removed by the principal component analysis
(PCA) technique (e.g. Li et al. 2005; Hao et al. 2005; Wang \& Wei
2008). Briefly, a library of pure stellar absorption-line spectra
(eigen-spectra) is built by applying the PCA method on the single
stellar population (SSP) models developed by Bruzual \& Charlot
(2003, hereafter BC03). The range of ages for these models is
between 1$\times$$10^{5}$ and 2$\times$$10^{10}$ yr. In our fitting
progress, several components are adopted and listed as follows: the
first seven eigen-spectra, an FeII complex template from BG92, and a
Galactic extinction curve (Cardelli et al. 1989). A $\chi^{2}$
minimizing algorithm is performed to determine the stellar
absorption spectrum for each galaxy over the wavelength range from
3700-6800 \AA, except for the regions around the strong emission
lines: H$\alpha$, H$\beta$, [NII]$\lambda$$\lambda$6548, 6584,
[OIII]$\lambda$$\lambda$4959, 5007, and [OII]$\lambda$3727. The left
column of Figure 1 shows the results of the AGN continuum/starlight
removal, the modeled FeII complex, and the isolated AGN
emission-line spectra.

The AGN emission-line spectra are then modeled by the multi-components
modeling procedure (SPECFIT task in IRAF package) as introduced by Kriss (1994).
Each emission line is modeled by a set of Gaussian profiles.
We model each of the H$\alpha$, H$\beta$ and [OIII] emission line profiles
by two Gaussians: one broad and one narrow components.
The width of
[OIII]$\lambda$5007$_n$ is set to be the same as that of
H$\beta_n$. Both intensity ratios [\ion{O}{3}]$\lambda$5007/$\lambda$4959
and [\ion{N}{2}]$\lambda$6583/$\lambda$6548 are
fixed to be 3.0. The profile modelings for the
H$\alpha$ and H$\beta$ regions are illustrated in Figure 1 (middle and right columns).
The results of the line profile modelings are shown in Table 1. All the
uncertainties given in Table 1 are caused by the profile modelings.

%% In a manner similar to \objectname authors can provide links to dataset
%% hosted at participating data centers via the \dataset{} command.  The
%% second curly bracket argument is printed in the text while the first
%% parentheses argument serves as the valid data set identifier.  Large
%% lists of data set are best provided in a table (see Table 3 for an example).
%% Valid data set identifiers should be obtained from the data center that
%% is currently hosting the data.
%%
%% Note that AASTeX interprets everything between the curly braces in the
%% macro as regular text, so any special characters, e.g. "#" or "_," must be
%% preceded by a backslash. Otherwise, you will get a LaTeX error when you
%% compile your manuscript.  Special characters do not
%% need to be escaped in the optional, square-bracket argument.

\section {RESULTS}

\subsection {Star-forming origin for narrow emission lines}
Basing upon a set of line ratios, such as [OIII]/H$\beta$,
[NII]/H$\alpha$, [SII]/H$\alpha$, and [OI]/H$\alpha$, the BPT
diagrams are commonly used as a powerful tool to determine the
dominant energy source in the emission-line galaxies. Comparing with
star-forming galaxies, AGNs (Seyferts and LINERs) have larger line
ratios because of their stronger and harder ionizing field. The
theoretical demarcation lines determine the lower limit of AGNs
(Seyfert and LINERs) were proposed by Kewley et al. (2001) according
to their theoretical stellar photoionization models. Using the large
spectra database provided by SDSS, Kauffmann et al. (2003) proposed
an empirical demarcation line separating star-forming galaxies in
the [OIII]/H$\beta$ vs. [NII]/H$\alpha$ diagram.

Because the Balmer line profiles are modeled by a combination of a
broad and a narrow components in the three objects, it is therefore
necessary to re-examine their locations on the BPT diagrams in the
current study. The three objects are plotted in the BPT diagrams in
Figure 2. The four panels represent the four different pairs of line
ratios. It is clear that all the three objects are below the
theoretical demarcation lines (see Panel A, B and C) proposed in
Kewley et al. (2001). In particular, SDSS J091053+333008 is
obviously located in the HII region in the [\ion{O}{3}]/H$\beta$ vs.
[\ion{N}{2}]/H$\alpha$ diagram, i.e, far below the empirical
classification line suggested by Kauffmann et al. (2003). In the
same diagram, the other two objects are located marginally above the
empirical demarcation line. Since the objects are extracted from
star-forming galaxies from the MPA/JHU catalog, the reason why they
are not located in the HII region is that we have removed the broad
component of H$\alpha$ and H$\beta$. As a result, the ratio of
[OIII]/H$\beta$ and [NII]/H$\alpha$ might be a little different from
Kauffmann et al. (2003).

As an additional test, we plot the three objects on the
[\ion{O}{3}]/[\ion{O}{2}] vs. [\ion{O}{1}]/H$\alpha$ diagram in
Panel D in Figure 2. The line ratio of [\ion{O}{3}]/[\ion{O}{2}]
is sensitive to the ionization parameter
of the gas (Kewley et al. 2006). To obtain the line ratio,
we need to correct the intrinsic dust extinction on the flux of
[\ion{O}{3}] and [\ion{O}{2}]. For each object, the
extinction is estimated from the observed Balmer decrement for
narrow components of H$\alpha$ and H$\beta$,
assuming the Case B recombination. The predicted
H$\alpha$/H$\beta$ ratio is 2.86 for \ion{H}{2} region in the
condition of gas density 100$\rm{cm^{-3}}$ and temperature
$10^{4}$K (Osterbrock 1989). The derived
$A_v$ values [=$R_v$$\times$E(B-V)] are listed in Table 1 for the
three objects. As shown in the diagram, all of our three objects
are located in the region occupied by star-forming galaxies.

In summary, using the widely used diagnostic tools,
we draw a conclusion that the narrow emission lines in the three objects
are mainly powered by
the ionizing field from hot stars rather than AGNs.

After demonstrating that the narrow emission lines are mainly
powered by star formation activity, we attempt to estimate current
star formation rate for the three objects in terms of the
[\ion{O}{2}]$\lambda$3727 emission. [\ion{O}{2}]$\lambda$3727 is an
empirical indicator to estimate the current star formation rate
(SFR) for emission-line galaxies (e.g. Gallagher et al. 1989;
Kennicutt 1998; Kewley et al. 2004). We calculate the SFR according
to the equations in Kewley et al. (2004):

\begin{equation}
\mathrm{SFR([OII])} = \frac{7.9\times10^{-42} \mathrm{L([OII])(ergs
s^{-1})}}{(-1.75)[\mathrm{log(O/H)}+12]+16.73}\
M_{\odot}\rm{yr^{-1}}
\end{equation}
The metallicity is estimated according to the empirical calibration
using the $R_{23}$ indicator (Kewley et al. 2004), where
$R_{23}=\mathrm{\log([OII]\lambda3727+[OIII]\lambda5007)/H\beta}$.
Again the [\ion{O}{2}] luminosity is corrected for the intrinsic
extinction derived from the Balmer decrement. The calculated current
SFRs of the three objects are listed in Table 1. It turns out that
all of them have small SFR$\sim1\rm{M_\odot\ yr^{-1}}$, which
indicates a suppressed star formation activity likely due to the AGN
feed back (see below for discussion of AGN properties).

\subsection {Broad emission lines: an indicative of AGNs}

The AGN properties are examined in this section. The detection of
broad Balmer emission lines is an indicative of the existence of AGN
in the center of galaxy (see Section 4 for the discussion of other
origins). Our spectral measurements indicate that all of the three
objects show AGN line emission with relatively narrow H$\beta$
profiles and large \ion{Fe}{2} ratios, which are the typical
features of Narrow-line Seyfert 1 Galaxies (NLS1s, e.g., Osterbrock
\& Pogge 1985). The measured FWHMs of H$\beta$ broad emission lines
are around $2000\mathrm{km\ s^{-1}}$, and the \ion{Fe}{2} ratios
$\mathrm{RFe}\sim2$ (except SDSS\,153002-020415), where the RFe is
defined as the flux ratio of optical
\ion{Fe}{2}/H$\beta_\mathrm{B}$. The flux of the \ion{Fe}{2} blends
is measured between the rest frame wavelength 4434 and 4684\AA.
Especially, the object SDSS\,J121837+091324 has already been
included in the $\sim$2000 NLS1 galaxies sample from SDSS by Zhou et
al. (2006).

The co-evolution of AGN and its host galaxy implies that young AGNs
are associated with young stellar population and intensive star
formation activity. Mathur (2000) proposed that NLS1s with small
black hole mass ($M_{\mathrm{BH}}$) and high Eddington ratio
($L/L_{\mathrm{Edd}}$) are at early stage of evolution of AGN. The
evident broad emission lines allow us to study the black hole
accretion directly in the three objects. The black hole mass
($M_{\mathrm{BH}}$) and Eddington ratio ($L/L_{\mathrm{Edd}}$) are
two basic physical parameters of AGNs (e.g. BG92; Boroson 2002). We
refer the readers to McGill et al. (2008) for a summary of the
existing formula used to estimate $M_{\mathrm{BH}}$ basing upon
``single-epoch'' spectrum. These methods are based on the great
progress in the calibration of the $R_\mathrm{BLR}-L$ relationship
(e.g., Kaspi et al. 2007 and references therein). In the current
study, we estimate the $M_{\mathrm{BH}}$ for the three objects
according to their broad $\rm{H\alpha}$ and $\rm{H\beta}$ emission
lines (Greene \& Ho 2005):

\begin{equation}
M_{\rm{BH}}=2\times10^{6}\bigg(\frac{L_{\rm{H\alpha}}}{10^{42}\
\rm{ergs\ s^{-1}}}\bigg)^{0.55}
\bigg(\frac{\rm{FWHM(H\alpha)}}{1000\ \rm{km\
s^{-1}}}\bigg)^{2.06}M_{\odot}
\end{equation}
\begin{equation}
M_{\rm{BH}}=3.6\times10^{6}\bigg(\frac{L_{\rm{H\beta}}}{10^{42}\
\rm{ergs\ s^{-1}}}\bigg)^{0.56} \bigg(\frac{\rm{FWHM(H\beta)}}{1000\
\rm{km\ s^{-1}}}\bigg)^{2}M_{\odot}
\end{equation}

$\rm{L_{bol}}$/$\rm{L_{Edd}}$ is estimated by simply assuming the
bolometric luminosity of quasar is proportional to the luminosity at
rest-frame wavelength 5100\AA: $L_{\rm{bol}}=9\lambda L_\lambda$(5100\AA)
(e.g., Kaspi et al. 2000), where
\begin{equation}
L_{5100}=2.4\times10^{43}\bigg(\frac{L_{\rm{H}\alpha}}{10^{42}\
\rm{erg\ s^{-1}}}\bigg)^{0.86}\ \rm{erg\ s^{-1}}
\end{equation}

For each object, both the luminosities of H$\alpha$ and H$\beta$ are
corrected for local extinction according to their observed Balmer decrements.
The estimated
$\rm{M_{BH}}$ from $\rm{H\alpha}$ and $\rm{H\beta}$ are listed in
Table 1, and they are highly consistent with each other within a factor
of two for all the three galaxies, which is less than the typical
absolute uncertainties of the mass scaling relationships (Vestergaard \& Peterson 2006).
It is clear that all the three objects show low
$\rm{M_{BH}}\sim10^{6}$$\rm{M_{\odot}}$ and high $L/L_{\mathrm{Edd}}\sim0.2$,
which implies that the three objects have an NLS1-like nucleus in their galaxy
centers.

\section{Discussion}

The spectral analysis given above clearly reveals that the emission lines
of the three objects consist of two components with distinct origins.
The narrow emission lines are mainly powered by star formation activity, and the
broad emission lines by central AGN. In addition, the AGNs show
small $M_{\mathrm{BH}}$, large $L/L_{\mathrm{Edd}}$, and strong
\ion{Fe}{2} emission, which is typical of NLS1 galaxies (e.g., Zhou et al. 2006).

Except for the AGN's contribution, other mechanisms may be
responsible for the broad line emission in the spectra of these
objects, including stellar winds and supernova remnants (e.g. Izotov
et al. 2007). Both of these mechanisms can produce broad emission
lines with FWHM $\geq$ 1000 km $\rm{s^{-1}}$. However, the
luminosity of H$\alpha$ broad components for our three objects are
$\sim$ $10^{42}$ erg $\rm{s^{-1}}$, far exceeding the typical
luminosity of broad H$\alpha$ of stellar winds from Wolf-Rayet or
luminous blue variable stars ($10^{36}$-$10^{40}$ erg
$\rm{s^{-1}}$). Moreover, if the broad component is caused by
supernova events, the magnitude obtained from the spectrum is
expected to greatly differ from the photometric magnitude, because
the brightness of supernova must decay at the time interval between
the two observations of about a year. We obtain the SDSS $r'$ band
magnitudes from the observed spectra by convolving each spectrum
with the $r'$ band transmission curve, and compare them with the
corresponding fiber magnitude obtained in the SDSS The magnitude
difference is found to be less than 1 mag within the interval of
about a year for these three objects. As a result, we can eliminate
the possibility of Type IIp supernova events.

We propose that the three objects are at the transition stage from
the starburst-dominated phase to AGN-dominated phase, which is supported
by the theoretical evolutionary scenario given by the numerical simulations
of mergers of two gas rich galaxies. Hopkins et al. (2005a, 2005b, and
references therein) proposed a scenario that the central
black hole of AGN is buried by gas or dust in the young AGN phase.
In their analysis, when starbursts are triggered by the nuclear
inflow and feed black hole growth, the black hole is heavily
obscured by surrounding gas or dust. As the black hole grows, the
black hole feedback will blow the gas or dust away in a powerful
wind, and the luminous AGN will be seen at that time. A case for the
model is NGC 6240, of which the X-ray data reveal that there are two
AGNs, while the optical data cannot resolved them, possibly for the
reason that the AGNs are in highly dust-enshrouded environments
(e.g. Komossa et al. 2003; Gressen et al. 2004; Max et al. 2005).
We believe that the next generation hard X-ray missions with enhanced
capability (in sensitivity and imaging), such as NuSTAR, Simbol-X and NeXT,
would be helpful in identifying and studying the starburst obscured AGNs.

Finally, we need to understand why the emission from AGN NLR is so
weak that the narrow emission is overwhelmed by the contribution
from \ion{H}{2} region. Netzer et al. (2004, and see also in Netzer
et al. 2006) proposed that the NLR would be dynamically unbounded if
the AGN luminosity is high enough. The distance of the NRL, if it
exists, from the central black hole is estimated to be not far less
than 1kpc for the three objects according to the relationship
$R_{\mathrm{NLR}}=1.15L_{\mathrm{H\beta,42}}^{0.49}\ \mathrm{kpc}$
(Netzer et al. 2004), where
$L_{\mathrm{H\beta},42}=L_{\mathrm{H\beta}}/10^{42}\ \mathrm{erg\
s^{-1}}$. At this distance, the escape velocity for a spherical
galaxy roughly equals to
$290M_{*,11}^{1/2}R_{\mathrm{10kpc}}^{-1/2}\ \mathrm{km\ s^{-1}}$,
where $M_{*,11}$ is the stellar mass in units of $10^{11}M_\odot$,
and $R_{\mathrm{10kpc}}$ the radial distance in units of 10kpc. The
stellar mass could be inferred from the the estimated mass of the
SMBH according to the well-established black hole and bulge mass
relation $\langle M_\mathrm{BH}/M_*\rangle\sim0.002$ (McLure \&
Dunlop 2004), which results in an escape velocity only $\sim100\
\mathrm{km\ s^{-1}}$. This scenario is supported by recent
observation study of extended emission-line regions (EELRs) around
low-z QSOs. Husemann et al. (2008) found that the presence of EELRs
are preferentially detected in QSOs with large black hole masses.

The lack of typical NLR in the three objects leads us to suspect
that the NLR might evolve with AGN and its host galaxies. Netzer et
al. (2004) proposed a scenario of ``starforming NLRs'' in which the
compact NLR gas in kpc scale is caused by the star formation
activity. An alternative possible scenario proposed here is that the
the formation of NLR is the consequence of the feedback of central
AGNs. As discussed above, a powerful feedback wind is required for
AGNs to dispel the surrounding gas and dust material to appear as a
luminous AGN at about 1 Gyr since the merger (Hopkins et al. 2005a).
This implies the typical NLR would not emerge until a majority of
surrounding material has been dispelled. Another implication for the
wind scenario is that the NLR kinematics is dominated by radial
outflow. In fact, spectroscopic studies with high spatial resolution
of NLR of a few nearby Seyfert 2 galaxies indicate that radial
velocity fields can be modeled by biconical outflows where the NLR
gas accelerates outward from the nucleus to a turnover point at a
distance of $\sim$100pc from the nucleus and subsequently appears to
decelerate to the systemic velocity at a distance of $\sim$300pc
(e.g., Hutchings et al. 1998; Kaiser et al. 2000; Nelson et al.
2000; Crenshaw et al. 2000; Crenshaw \& Kraemer 2000; Ruiz et al.
2001).

\section{Conclusions}

By studying the optical spectra of the three objects: SDSS
J091053+333008, SDSS J121837+091324, and SDSS J153002-020415, we
find that all of them are located in the HII region in the BPT
diagnostic diagram. However, all of their spectra display
characteristic features of AGNs. All of them have large Eddington
ratio, small black hole mass, and small SFR. We propose that the
dust-buried-AGN (Hopkins et al. 2005a, b) can explain the feature.

\begin{acknowledgements}

We would thank the anonymous referee for constructive comments that
improved the paper. We are grateful to Todd A. Boroson and Richard
F. Green for providing us the FeII template. This work was supported
by the National Science Foundation of China (grant 10803008 and
10873017) and National Basic Research Program of China. The SDSS
archive data is created and distributed by the Alfred P. Sloan
Foundation. This research has made use of the NASA/IPAC
Extragalactic Database, which is operated by JPL, Caltech, under
contact with the NASA.

\end{acknowledgements}

\clearpage
\begin{figure}
\plotone{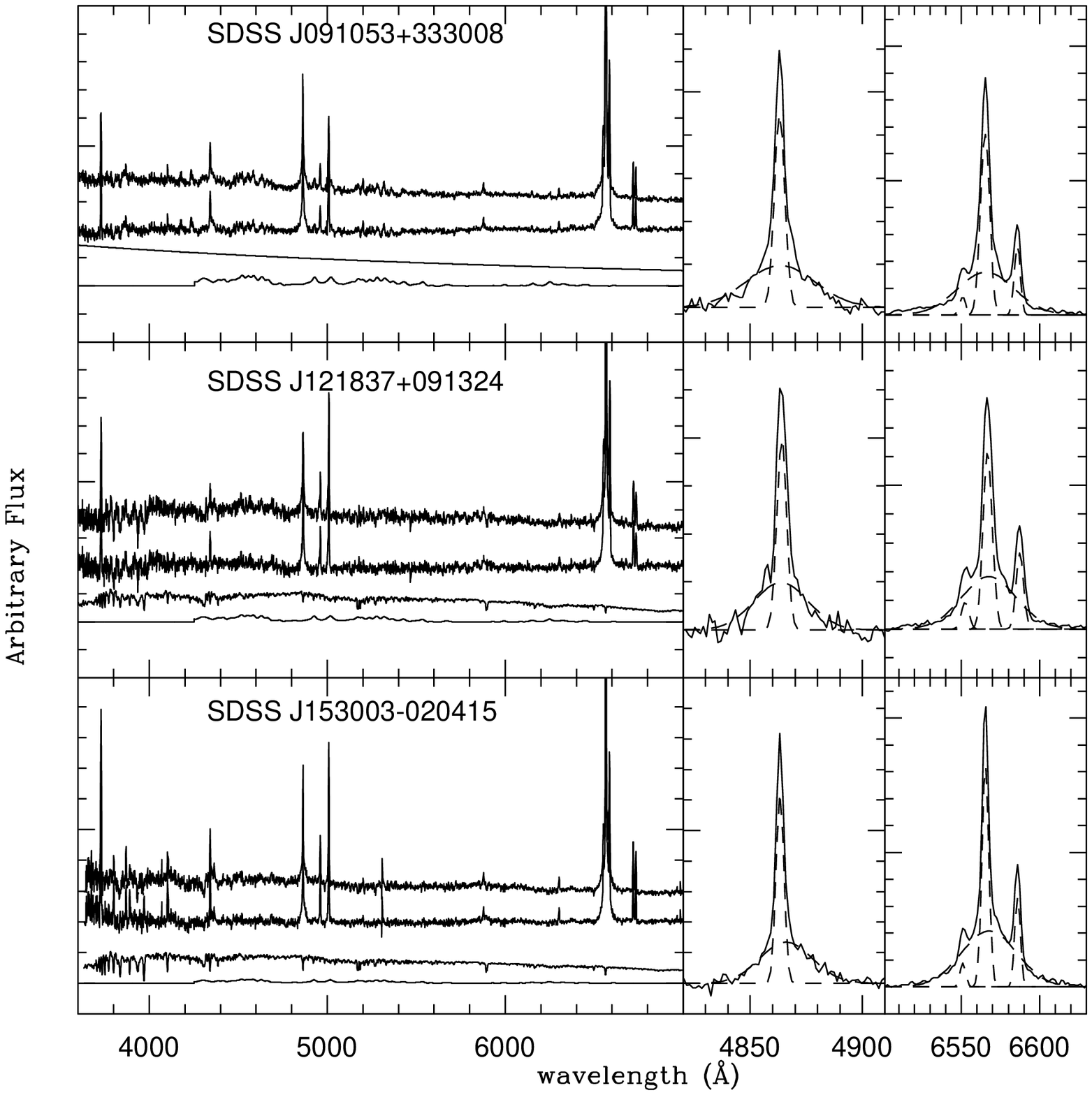} \caption{left: Detailed modeling and subtractions
of the starlight components in the spectra of the three objects. For
SDSS J091053+333008, we plot the observed spectrum, emission-line
spectrum, a power-law continuum, and FeII blends, from top to
bottom. For the other two objects, we plot the emission-line
spectrum, observed spectrum, modeled starlight spectrum, and FeII
blends. The spectra are vertically shifted by arbitrary amounts for
visibility. Middle: Modelings of the two Gaussion profiles for the
H$\beta$ region. Right: Modelings of a set of Gaussion profiles for
the H$\alpha$ region.\label{fig1}}
\end{figure}

\clearpage
\begin{figure}
\plotone{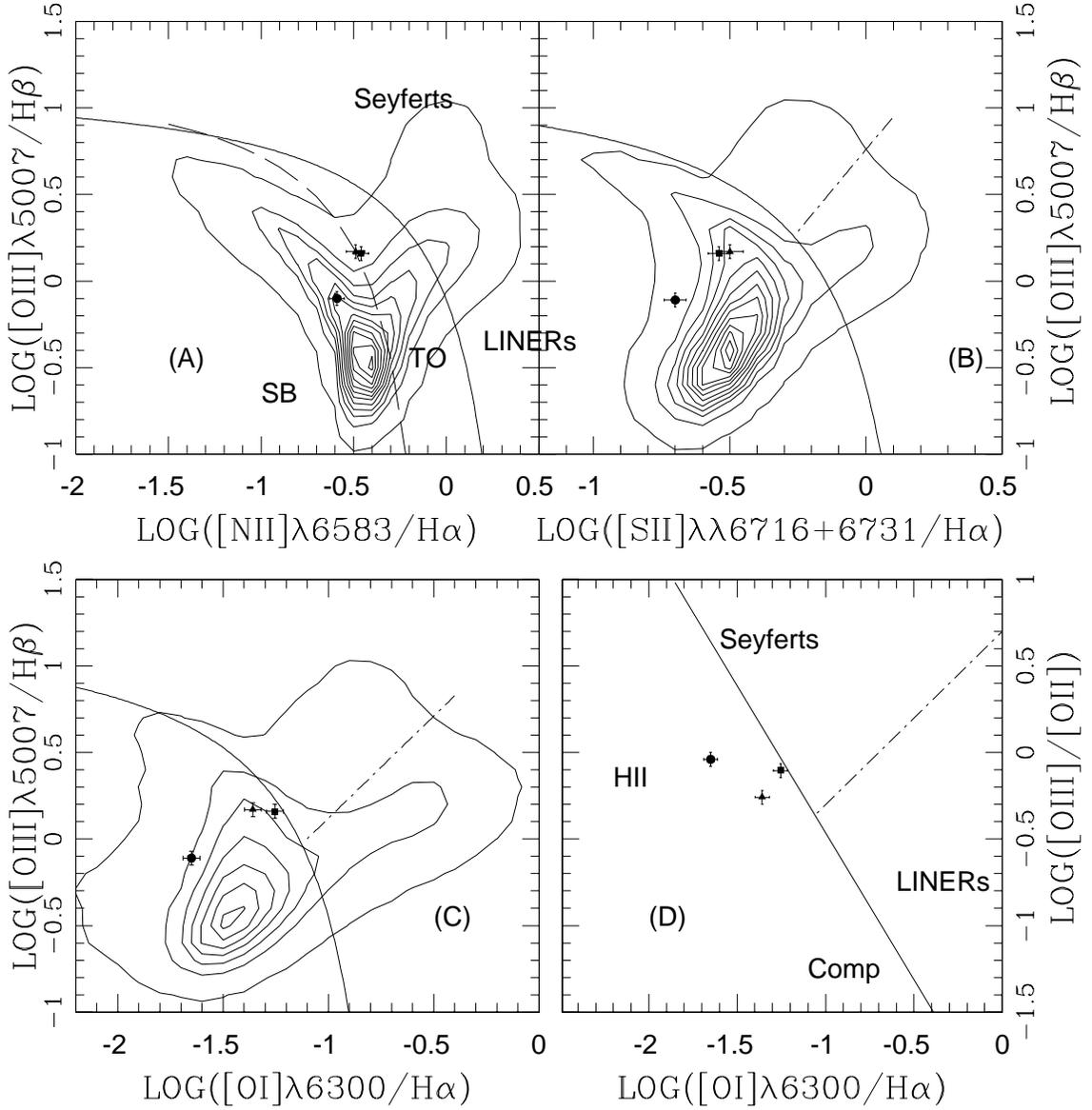} \caption{The four BPT diagnostic diagrams for our
three galaxies. The theoretical lines separating AGNs from
star-forming galaxies proposed by Kewley et al. (2001) are shown by
the solid lines in all four panels, while the empirical lines
proposed by Kauffmann et al. (2003) by the dashed line in Panel A.
The dot-dashed lines in Panel B, C, D separating LINERS from
Seyferts are proposed by Kewley et al. (2006). The circles are for
SDSS\,J091053+333008, the squares are for SDSS\,J121837+091324, and
the triangles are for SDSS\,J153002-020415. All of our three
galaxies are located in the star-forming region, which is proposed
by Kewley et al. (2006). \label{fig2}}
\end{figure}

\begin{table}
\begin{center}
\caption {Properties of the emission lines of the three objects,
Black hole masses, Eddington Ratios and SFR.}

\begin{tabular}{cccccc}
\tableline\tableline
Lines & SDSS J091053+333008 & SDSS
J121837+091324 & SDSS J153002-020415 \\
\tableline
H$\beta_{\rm{N}}$ & 25.8$\pm$1.3   &  11.1$\pm$0.7 & 27.5$\pm$2.0 \\
H$\beta_{\rm{B}}$ & 37.3$\pm$1.7  &  14.8$\pm$0.9 & 39.5$\pm$2.8 \\
$[\rm{OIII}]\lambda$5007 & 20.3$\pm$0.7  &  16.1$\pm$0.4 & 41.1$\pm$0.5 \\
H$\alpha_{\rm{N}}$ & 99.5$\pm$3.1 & 45.5$\pm$1.5 & 93.0$\pm$4.1 \\
H$\alpha_{\rm{B}}$ & 148.1$\pm$4.0 & 79.0$\pm$2.2 & 185.8$\pm$6.4 \\
$[\rm{NII}]\lambda6583$ & 25.3$\pm$1.2 & 15.9$\pm$0.7 & 30.2$\pm$2.3 \\
$[\rm{SII}]\lambda$6716 & 10.0$\pm$0.6 & 7.8$\pm$0.4 & 17.4$\pm$1.4 \\
$[\rm{SII}]\lambda$6731 & 9.5$\pm$0.6 & 5.2$\pm$0.4 & 11.8$\pm$1.2 \\
$[\rm{OI}]\lambda$6300  & 2.2$\pm$0.5 & 2.4$\pm$0.5 & 4.1$\pm$0.9 \\
$[\rm{OII}]\lambda$3727  & 17.4$\pm$1.0 & 14.8$\pm$0.7 & 65.5$\pm$3.5 \\
Rfe & 1.7$\pm$0.1 & 2.9$\pm$0.2 & 0.1$\pm$0.0 \\
FWHM(H$\beta_{\rm{B}}$) & 2.3$\pm$0.2 & 1.7$\pm$0.2 & 1.7$\pm$0.2
\\ \tableline
$A_v$  & 0.70$\pm$0.04 & 0.88$\pm$0.04 & 0.35$\pm$0.02\\
$\rm{M_{BH}}$(H$\alpha$)/$10^{6}$$\rm{M_{\odot}}$ & 7.9$\pm$2.0 & 3.2$\pm$0.7 & 2.8$\pm$0.6\\
$\rm{M_{BH}}$(H$\beta$)/$10^{6}$$\rm{M_{\odot}}$ & 9.9$\pm$2.4 & 2.3$\pm$0.7 & 1.7$\pm$0.6\\
$\rm{L_{bol}}$/$\rm{L_{Edd}}$ & 0.21$\pm$0.04 & 0.16$\pm$0.03 & 0.13$\pm$0.03 \\
SFR & 2.4$\pm$0.1 & 0.8$\pm$0.0 & 0.6$\pm$0.0\\
 \tableline \tableline
\end{tabular}
%% Any table notes must follow the \end{tabular} command.
Notes: The flux of each component is in the unit of $10^{-16}
\rm{ergs\ s^{-1}\ cm^{-2}}$. The FWHM(H$\beta$) is in the unit of
10$^{3}$km\ $\rm{s^{-1}}$. And the SFR is in the unit of
$\rm{M_{\odot}}$\ $\rm{yr^{-1}}$
\end{center}
\end{table}

%% The following command ends your manuscript. LaTeX will ignore any text


\begin{thebibliography}{}
\bibitem[Adelman-McCarthy et al. 2006]{am06} Adelman-McCarthy, J. K., et al. 2006, ApJS, 162, 38
\bibitem[Balogh et al. 1999] {ba99} Balogh, M. L., et al., 1999, \apj, 527, 54
\bibitem[Boroson \& Green 1992]{bg92} Boroson, T. A., \& Green, R. F., 1992, \apjs, 80, 109
\bibitem[Boroson 2002] {bo02} Boroson, T. A., 2002, \apj, 565, 78
\bibitem[Boroson 2005] {bo05} Boroson, T. A., 2005, \apj, 130, 381
\bibitem[Brotherton et al. 1999]{bo00} Brotherton, M. S., van Breugel, Wil., Stanford, S. A., et al., 1999, \apj, 520, L87
\bibitem[Bruzual \& Charlot 2003]{bc03} Bruzual. G., \& Charlot. S., 2003, \mnras, 344, 1000
\bibitem[Canalizo et al. 2006]{ca06} Canalizo, G., et al., 2006, NewAR, 50, 650
\bibitem[Cardelli et al. 1989]{ca89} Cardelli, J. A., Clayton, G. C., \& Mathis, J. S., 1989, \apj, 345, 245
\bibitem[Crenshaw et al. 2000]{cr00} Crenshaw, D. M., et al. 2000, \aj, 120, 1731
\bibitem[Crenshaw \& Kraemer 2000]{ck00} Crenshaw, D. M., \& Kraemer, S. B. 2000, \apj, 532, L101
\bibitem[Di Matteo et al. 2005]{DM05} Di Matteo, T., et al., 2005, nature, 433, 604
\bibitem[Ferrarese et al. 2006]{fe06} Ferrarese, L., et al. 2006, \apj, 644, L21
\bibitem[Gallagher et al. 1989]{ga89} Gallagher, J. S., Hunter, D. A., \& Bushouse, H., 1989, \aj, 97, 100
\bibitem[Goncalves et al. 1999] {go99} Goncalves, A. C., Veron-Cetty, M. P., \& Veron, P., 1999, A\&AS, 135, 437
\bibitem[Gonzalez Delgado 2002]{go02} Gonzalez Delgado, R., 2002, ASPC, 258, 101
\bibitem[Goto 2006]{go06} Goto, T., 2006, \mnras, 369, 1765
\bibitem[Greene \& Ho 2005] {gh05} Greene, J. E., \& Ho, L. C., 2005, \apj, 630, 122
\bibitem[Greene \& Ho 2006] {gh06} Greene, J. E., \& Ho, L. C., 2006, \apj, 641, L21
\bibitem[Gressen et al. 2004] {gr04} Gressen, J., et al., 2004, \aj, 127, 75
\bibitem[Hao et al. 2005]{ha05} Hao, L., Strauss, M. A., Tremonti, C. A., et al. 2005, \aj, 129, 1783
\bibitem[Heckman et al. 2004]{he04} Heckman, T. M., Kauffmann, G., Brinchmann, J., et al., 2004, \apj, 613, 109
\bibitem[Hopkins et al. 2005a]{ho05a} Hopkins, P., Hernquist, L., Martini, P., et al. 2005, \apj, 625, L71
\bibitem[Hopkins et al. 2005b]{ho05b} Hopkins, P., Hernquist, L., Cox, T., et al. 2005, \apj, 630, 705
\bibitem[Hu et al. 2008]{hu08} Hu, C., et al. 2008, \apj, 687, 78
\bibitem[Husemann et al. 2008[{hu08} Husemann, B., Wisotzki, L., Sanchez, S. F. \& Jahnke, K., 2008, A$\&$A, 488, 145
\bibitem[Hutchings et al. 1998]{hu98} Hutchings, J. B., et al. 1998, \apj, 492, L115
\bibitem[Kaiser et al. 2000]{ka00} Kaiser, M. E., et al. 2000, \apj, 528, 260
\bibitem[Kaspi et al. 2000] {ka00} Kaspi, S., et al. 2000, \apj, 539,  L13
\bibitem[Kaspi et al. 2007] {ka07} Kaspi, S., et al. 2007, \apj, 659, 997
\bibitem[Izotov et al. 2007]{izo07} Izotov, Y. I., Thuan, T. X., \& Guseva, N. G., 2007, \apj, 671, 1297
\bibitem[Kauffmann et al. 2003]{ka03} Kauffmann, G., Heckman, T. M., Tremonti, C., et al., 2003, \mnras, 2003, 346, 1055
\bibitem[Kennicutt 1998]{ke98} Kennicutt, R. C., 1998, ARA\&A, 36, 189
\bibitem[Kewley et al. 2001]{ke01} Kewley, L. J., et al. 2001, \apj, 556, 121
\bibitem[Kewley et al. 2004]{ke04} Kewley, L. J., Geller, M. J., \& Jansen, R. A., 2004, \aj, 127, 2002
\bibitem[Kewley et al. 2006]{ke06} Kewley, L. J., Groves, B., Kauffmann, G., Heckman, T., 2006, \mnras, 372, 961
\bibitem[Komossa et al. 2003] {ko03} Komossa, S., Burwitz, V., Hasinger, G., et al., 2003. \apj, 582, 15
\bibitem[Komossa et al. 2008] {ko08} Komossa, S., Xu, D., Zhou, H., et al., 2008. \apj, 680,
926
\bibitem[Kriss 1994] {kr94} Kriss, G., 1994, in ASP Conf. Ser. 61, Astronomical Data Analysis Software and Systems III, ed. D. R. Crabtree, R. J. Hanisch, \& J. Bames (San Francisco: ASP), 437
\bibitem[Li et al. 2005]{li05} Li, C., Wang, T. G., Zhou, H. Y., et al. 2005, \aj, 129, 669
\bibitem[Magorrian et al. 1998]{ma98} Magorrian, J., Tremaine, S., Richstone, D., et al., 1998, \aj, 115, 2285
\bibitem[Mathur 2000]{ma00} Mathur, S. 2000, MNRAS, 314, L17
\bibitem[McGill et al. 2008]{mg08} McGill, K. L., et al. 2008, \apj, 673, 703
\bibitem[McLure \& Dunlop 2004]{md04} McLure, R. J., \& Dunlop, J. S., 2004, \mnras, 352, 1390
\bibitem[Max et al. 2005] {ma05} Max, C. E., et al., 2005, \apj, 621, 738
\bibitem[Nelson 2000]{ne00} Nelson, C. H. 2000, \apj, 544, L91
\bibitem[Netzer et al. 2004]{ne04} Netzer, H., et al. 2004, \apj, 614, 558
\bibitem[Netzer et al. 2006]{ne06} Netzer, H., et al. 2006, A$\&$A, 453, 525
\bibitem[Osterbrock \& Pogge 1985]{op85} Osterbrock, D. E., \& Pogge, R. W. 1985, \apj, 297, 166
\bibitem[Osterbrock 1989]{os89} Osterbrock, D. E., 1989, Astrophysics of Gaseous Nebulae and Active Galactic Nuclei, Mill Valley  CA: University Science Books
\bibitem[Riechers et al. 2008]{ri08} Riechers, D. A., Walter, F., Carilli, C. L., Bertoldi, F., Momjian, E., 2008, \apj, 686, L9
\bibitem[Schlegel et al. 1998]{sc98} Schlegel, D. J., Finkbeiner, D. P., \& Davis, M. 1998, ApJ, 500, 525
\bibitem[Shields et al. 2003]{sh03} Shields, G., Gebhardt, K., Salviander, S., et al., 2003, \apj, 583, 124
\bibitem[Springel et al. 2005]{sp05} Springel, V., et al. 2005, nature, 435, 629
\bibitem[Tremaine et al. 2002]{tr02} Tremaine, S., et al. 2002, ApJ, 574, 740
\bibitem[Vestergaard \& Peterson. 2006] {vp06} Vestergaard, M., \& Peterson, B., 2006, \apj, 641, 689
\bibitem[Walter et al. 2004]{wa04} Walter, F., Carilli, C., \& Bertoldi, F., et al. 2004, 615, L17
\bibitem[Wang et al. 2006]{wa06} Wang, J., Wei, J. Y., \& He, X. T., 2006, \apj, 638, 106
\bibitem[Wang \& Wei, 2008] {ww08} Wang, J., \& Wei, J. Y., 2008, \apj, 679, 86
\bibitem[Weedman 1983] {we83} Weedman, D. W., 1983, \apj, 266, 479
\bibitem[Wild et al. 2007]{wi07} Wild, V., et al. 2007, \mnras, 381, 543
\bibitem[Worthey \& Ottaviani, 1997] {wo97} Worthey, G., \& Ottaviani, D. L., 1997, \apjs, 111, 377
\bibitem[Zhou, et al. 2005]{zh05} Zhou, H. Y., Wang, T. G., Dong, X. B., Wang, J., Lu, H., 2005, Mem, S. A. It 76, 93
\bibitem[Zhou, et al. 2006]{zh06} Zhou, H. Y., et al. 2006, ApJS, 166, 128

\end{thebibliography}
\end{document}